\documentclass{revtex4}
\usepackage{graphicx}
\def\aprle{\buildrel < \over {_{\sim}}}

\begin{document}
\bibliographystyle{revtex}
\begin{flushright}
	FERMILAB-Conf-01/306-T\\
\end{flushright}
%\title{SO(10) SUSY GUT Model with a $U(1) \times Z_2 \times Z_2$
%	Flavor Symmetry\footnotemark}
%\setcounter{footnote}{1}
%\footnotetext{Contribution submitted for the proceedings of 
%	the Snowmass Workshop, 30 June - 21 July 2001.}
\title{SO(10) SUSY GUT Model with a $U(1) \times Z_2 \times Z_2$
	Flavor Symmetry\footnote{Contribution submitted for the proceedings of 
	the Snowmass Workshop on the Future of Particle Physics, 
	30 June - 21 July 2001.}}
\author{Carl H. Albright}
\email[]{albright@fnal.gov}
\affiliation{Fermi National Accelerator Laboratory, P.O. Box 500, Batavia, IL 
60510\\
and\\
Department of Physics, Northern Illinois University, DeKalb, IL 60115}
\date{October 15, 2001}
\begin{abstract}
An SO(10) SUSY GUT model which leads to maximal atmospheric neutrino mixing 
and the LMA solar neutrino solution, developed in collaboration with S.M. Barr,
is briefly described.  Since the model is quantitatively 
predictive, it can be used to assess the need for a neutrino factory,
as shown in collaboration with S. Geer.
\end{abstract}
\maketitle

The GUT model in question \cite{ab} is based on the grand unified group 
$SO(10)$ with a $U(1) \times Z_2 \times Z_2$ flavor symmetry.  The model 
involves a minimum 
set of Higgs fields which solves the doublet-triplet splitting problem.  This 
requires just one ${\bf 45}_H$ whose VEV points in the $B-L$ direction, 
two pairs of ${\bf 16}_H,\ {\bf \overline{16}}_H$'s which stabilize the 
solution, along with several Higgs in the ${\bf 10}_H$ representations and
Higgs singlets \cite{br}.  The Higgs superpotential exhibits the
$U(1) \times Z_2 \times Z_2$ symmetry which is used for the 
flavor symmetry of the GUT model.  The combination of VEVs, 
$\langle {\bf 45_H}\rangle_{B-L},\ \langle 1({\bf 16_H})\rangle$ and 
$\ \langle 1({\bf \overline{16}_H})\rangle$ break $SO(10)$ to the Standard 
Model.  The electroweak VEVs arise from the combinations 
$v_u = \langle 5({\bf 10_H})\rangle$ and $v_d = \langle \overline{5}({\bf 10_H})
\rangle\cos \gamma + \langle \overline{5}({\bf 16'_H})\rangle \sin \gamma$,
while the combination orthogonal to $v_d$ gets massive at the GUT scale.
As such, Yukawa coupling unification can be achieved at the GUT scale with
$\tan \beta \sim 2 - 55$, depending upon the $\overline{5}({\bf 10_H}) -
\overline{5}({\bf 16_H})$ mixing present for the $v_d$ VEV.
In addition, matter superfields appear in the following representations:
${\bf 16_1},\ {\bf 16_2},\ {\bf 16_3};\ {\bf 16},\ {\bf \overline{16}},\ 
{\bf 16'},\ {\bf \overline{16'}}$, ${\bf 10_1},\ {\bf 10_2}$, and ${\bf 1}$'s,
where all but the ${\bf 16_i}\ (i = 1,2,3)$ get superheavy and are integrated
out.

The Dirac mass matrices for the up quarks, down quarks, neutrinos and charged 
leptons are found to be
\begin{equation}
\begin{array}{ll}
U = \left(\matrix{ \eta & 0 & 0 \cr
  0 & 0 & \epsilon/3 \cr 0 & - \epsilon/3 & 1\cr} \right)M_U,\
  & D = \left(\matrix{ 0 & \delta & \delta' e^{i\phi}\cr
  \delta & 0 & \sigma + \epsilon/3  \cr
  \delta' e^{i \phi} & - \epsilon/3 & 1\cr} \right)M_D, \\[0.3in]
N = \left(\matrix{ \eta & 0 & 0 \cr 0 & 0 & - \epsilon \cr
        0 & \epsilon & 1\cr} \right)M_U,\
  & L = \left(\matrix{ 0 & \delta & \delta' e^{i \phi} \cr
  \delta & 0 & -\epsilon \cr \delta' e^{i\phi} &
  \sigma + \epsilon & 1\cr} \right)M_D.\\
\end{array}\label{eq:P2-E1-albrighteq1}
\end{equation}
The above textures were obtained by imposing the Georgi-Jarlskog relations
\cite{gj} at $\Lambda_{GUT}$, $m^0_s \simeq m^0_\mu/3,\ m^0_d \simeq 3m^0_e$
with Yukawa coupling unification holding for $\tan \beta \sim 5$.
The matrix element contributions can be understood in terms of
Froggatt-Nielsen diagrams \cite{fn} as explained in \cite{ab}.  

All nine quark and charged lepton masses, plus the three CKM angles and CP
phase, are well-fitted with the eight input parameters 
\begin{equation}
\begin{array}{rlrl}
        M_U&\simeq 113\ {\rm GeV},&\qquad M_D&\simeq 1\ {\rm GeV},\\
        \sigma&=1.78,&\qquad \epsilon&=0.145,\\
        \delta&=0.0086,&\qquad \delta'&= 0.0079,\\
        \phi&= 126^\circ,&\qquad \eta&= 8 \times 10^{-6},\\
\end{array}\label{eq:P2-E1-albrighteq2}
\end{equation}
defined at the GUT scale to fit the low scale observables after evolution 
downward from $\Lambda_{GUT}$:
\begin{equation}
\begin{array}{rlrl}
           m_t(m_t) &= 165\ {\rm GeV},\quad & m_{\tau} &= 1.777\ {\rm GeV},
                \\[0.1in]
           m_u(1\ {\rm GeV}) &= 4.5\ {\rm MeV},\quad & m_\mu &= 105.7\ 
                {\rm MeV},\\[0.1in]
           V_{us} &= 0.220, \quad & m_e &= 0.511\ {\rm MeV},\\[0.1in]
           V_{cb} &= 0.0395, \quad & \delta_{CP} &= 64^\circ.\\
\end{array}
\label{eq:P2-E1-albrighteq3}
\end{equation}
These lead to the following predictions:
\begin{equation}
\begin{array}{rlrl}
           m_b(m_b) &= 4.25\ {\rm GeV},\quad & m_c(m_c) &= 1.23\ {\rm GeV},
                \\[0.1in]
           m_s(1\ {\rm GeV}) &= 148\ {\rm MeV},\quad & m_d(1\ {\rm MeV}) 
                &= 7.9\ {\rm MeV},\\[0.1in]
           |V_{ub}/V_{cb}| &= 0.080,\quad & \sin 2\beta &= 0.64.\\
\end{array}
\label{eq:P2-E1-albrighteq4}
\end{equation}
With no extra phases present, aside from the one appearing in the CKM mixing 
matrix, the vertex of the
CKM unitary triangle occurs near the center of the presently allowed region
with $\sin 2\beta \simeq 0.64$.  The Hermitian matrices $U^\dagger U,\
D^\dagger D$, and $N^\dagger N$ are diagonalized with small left-handed
rotations, while $L^\dagger L$ is diagonalized by a large left-handed rotation.
This accounts for the small value of $V_{cb} = (U^\dagger_U U_D)_{cb}$,
while $|U_{\mu 3}| = |(U^\dagger_L U_\nu)_{\mu 3}|$ will turn out to be 
large for any reasonable right-handed Majorana mass matrix, $M_R$ \cite{abb}.

The effective light neutrino mass matrix, $M_\nu$, is obtained from the seesaw 
mechanism once the right-handed Majorana mass matrix, $M_R$, is specified.  
While the large atmospheric neutrino mixing $\nu_\mu \leftrightarrow 
\nu_\tau$ arises primarily from the structure of the charged lepton mass 
matrix, the solar and atmospheric mixings are essentially decoupled in the 
model, so the structure of the right-handed Majorana mass matrix determines the
type of $\nu_e \leftrightarrow \nu_\mu,\ \nu_\tau$ solar neutrino mixing.
Any one of the recently favored four solar neutrino mixing solutions
can be obtained.  The LMA solution relevant to this discussion requires some
fine-tuning and a hierarchical structure, but this can also be explained in 
terms of Froggatt-Nielsen diagrams.  The most general form for the 
right-handed Majorana mass matrix considered in \cite{ab} is
\begin{equation}
          M_R = \left(\matrix{c^2 \eta^2 & -b\epsilon\eta & a\eta\cr
                -b\epsilon\eta & \epsilon^2 & -\epsilon\cr
                a\eta & -\epsilon & 1\cr}\right)\Lambda_R,\\
\label{eq:P2-E1-albrighteq5}
\end{equation}
where the parameters $\epsilon$ and $\eta$ are those introduced in 
Eq.(\ref{eq:P2-E1-albrighteq1}) for the Dirac sector.  With $a \neq b = c$, 
the structure
of $M_R$ arises from one Higgs singlet which induces a $\Delta L = 2$ 
transition and contributes to all nine matrix elements while, by virtue of 
its flavor charge assignment, a second Higgs singlet breaks lepton number but 
modifies only the 13 and 31 elements of $M_R$.

As a numerical example, with just three additional input parameters:
$a=1,\ b=c=2$ and $\Lambda_R = 2.4 \times 10^{14}$ GeV, where the latter is
used to scale $\Delta m^2_{32}$, 
\begin{equation}
	M_\nu = \left(\matrix{ 0 & -\epsilon & 0\cr 
                        -\epsilon & 0 & 2\epsilon\cr 0 & 2\epsilon & 1\cr}
                        \right)M^2_U/\Lambda_R\\
\label{eq:P2-E1-albrighteq6}
\end{equation}
leads to 
\begin{equation}
\begin{array}{ll}
          \multicolumn{2}{l}{m_1 = 5.6 \times 10^{-3},\quad m_2 = 9.8 \times 
                10^{-3},\quad m_3 = 57 \times 10^{-3}\ {\rm eV},}\\[0.05in]
          M_1 = M_2 = 2.8 \times 10^{8}\ {\rm GeV},\quad & M_3 = 2.5 
                \times 10^{14}\ {\rm GeV},\\[0.05in]
          \Delta m^2_{32} = 3.2 \times 10^{-3}\ {\rm eV^2},
                \quad &\sin^2 2\theta_{\rm atm} = 0.994,
                \\[0.05in]
          \Delta m^2_{21} = 6.5 \times 10^{-5}\ 
                {\rm eV^2},\quad &\sin^2 2\theta_{\rm sol} 
                = 0.88,\\[0.05in]
          U_{e3} = -0.01395-0.00085i,\quad &\sin^2 2\theta_{\rm reac} = 
		0.0008,\\[0.05in]
          J = 2.0 \times 10^{-4},\quad \delta_{CP} = -3.5^\circ,\quad & 
                \chi_1 = -0.2^\circ, \quad \chi_2 = 0.1^\circ,\\ 
\end{array}
\label{eq:P2-E1-albrighteq7}
\end{equation}
to be compared with the present S-K atmospheric data \cite{atm} 
and best-fit point in the LMA region \cite{sol}
\begin{equation}
\begin{array}{rlrl}
	\Delta m^2_{32} &\simeq 3.2 \times 10^{-3}\ {\rm eV^2},&
		\quad \sin^2 2\theta_{23}&= 1.0, (\geq 0.89 {\rm\ at\ 90\% 
		\ c.l.}),\\
	\Delta m^2_{21} &= 7 \times 10^{-5}\ {\rm eV^2},&
                \quad \sin^2 2\theta_{\rm sol} &= 0.87.\\
\label{eq:P2-E1-albrighteq8}
\end{array}
\end{equation}

\begin{figure}[h]
%\centering\leavevmode
\includegraphics[width=3in,angle=0]{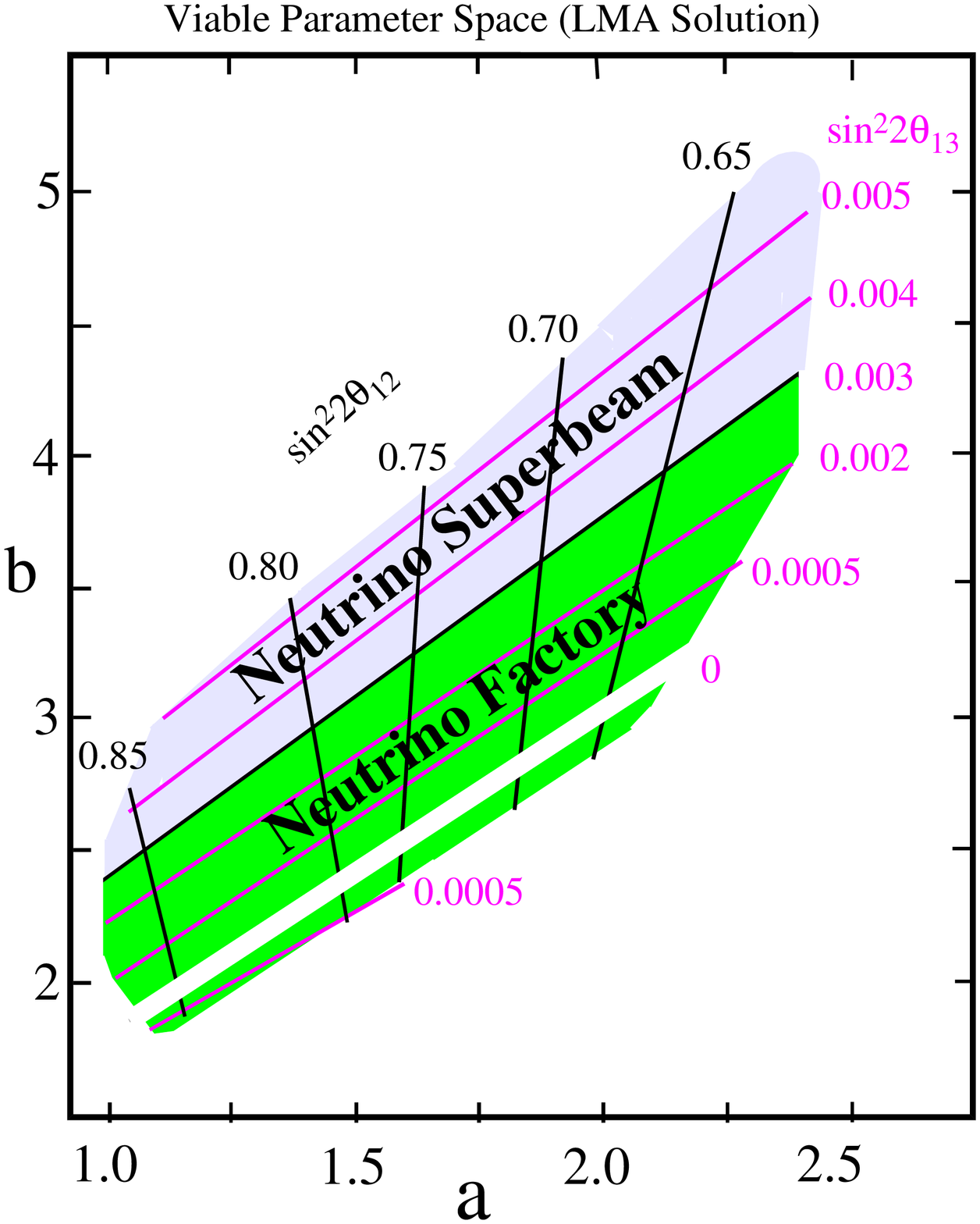}
\includegraphics[width=3in,angle=0]{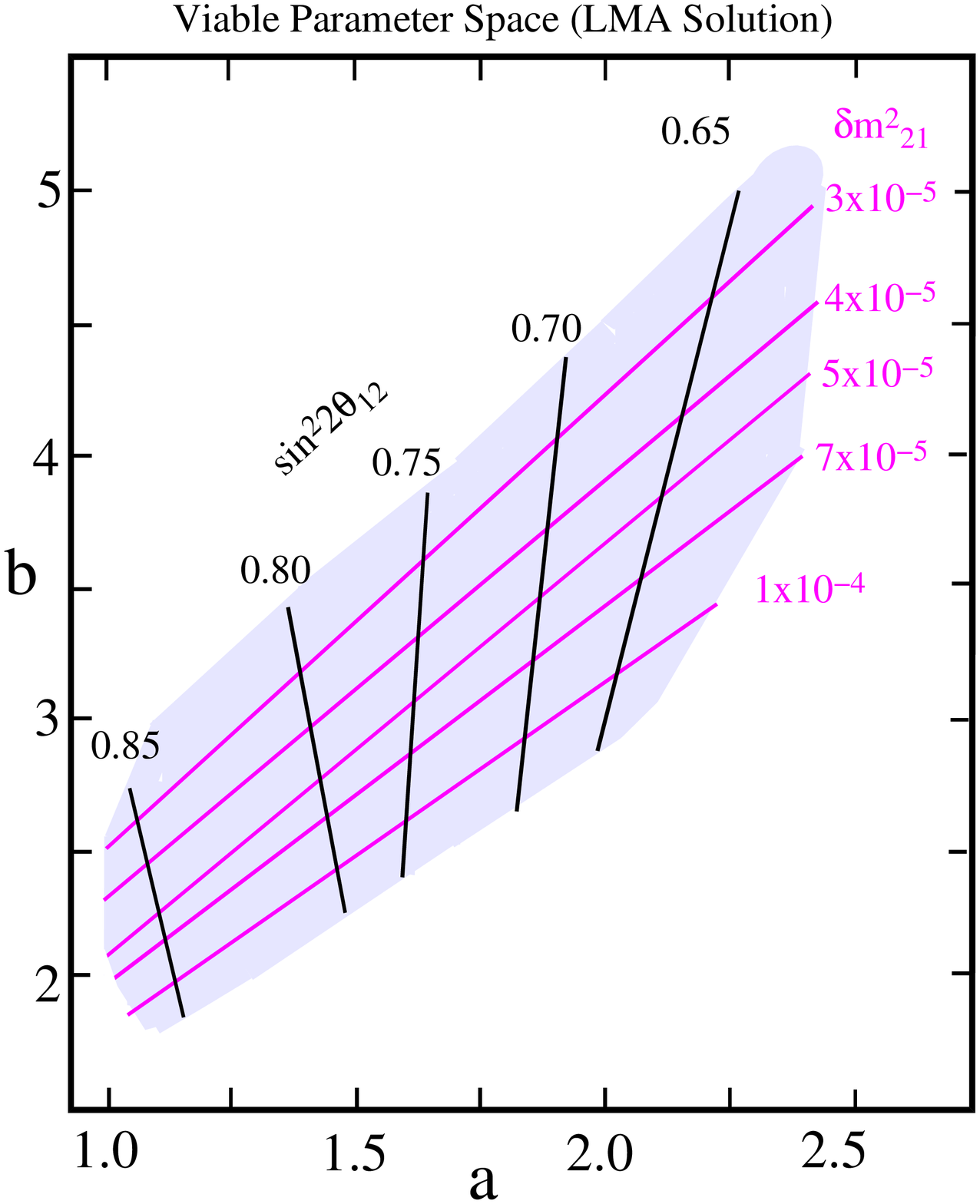}
\hspace*{0.25in}
\caption[]{The viable region of GUT parameter space consistent 
with the present bounds on the LMA MSW solution. Contours of 
constant $\sin^2 2\theta_{13}$ and lines of constant 
$\sin^2 2\theta_{12}$ are shown on the left.  The region above 
$\sin^2 2\theta_{13} = 0.003$ can be explored with 
Neutrino Superbeams, while the region below this can be explored 
with Neutrino Factories, down to $\sin^2 2\theta_{13} \sim 0.0001$.
Similarly, contours of constant $\Delta m^2_{21}$
along with lines of constant $\sin^2 2\theta_{12}$ are shown on the right.}
\label{fig:P2-E1-albrightfig1}
\end{figure}

In fact, the whole presently-allowed LMA region can be covered with 
\begin{equation}
	1.0 \aprle a \aprle 2.4,\quad 1.8 \aprle b=c \aprle 5.2.\\
\label{eq:P2-E1-albrighteq9}
\end{equation}
The viable region of GUT model parameter space consistent with the LMA solar
solution is shown in Fig.~\ref{fig:P2-E1-albrightfig1}.  Superimposed on the 
allowed region, Fig.~\ref{fig:P2-E1-albrightfig1} shows lines of constant 
$\sin^2 2\theta_{12}$ and contours of constant $\sin^2 2\theta_{13}$.
For the fully allowed parameter space, we see that $\sin^2 2\theta_{13} 
< 0.006$.  While a new generation of upgraded conventional neutrino 
beams is being considered \cite{superbeams},
and is expected to be able to probe the region $\sin^2 2\theta_{13} > 0.003$,
the ``superbeams'' will be able to measure the parameter $\theta_{13}$ if 
the solution lies in the upper part of the allowed $(a,b)$-plane indicated in 
the figure. A Neutrino Factory \cite{NF}, on the other hand, is expected to be 
able to probe down to values of $\sin^2 2\theta_{13}$ as low as $O(10^{-4})$, 
which will therefore cover the entire
allowed $(a,b)$-plane, except for a narrow band in which $\sin^2 2\theta_{13}
\to 0$ as $\sin^2 2\theta_{23}$ becomes maximal. 

Also in Fig.~\ref{fig:P2-E1-albrightfig1}, contours of constant 
$\Delta m^2_{21}$ are displayed within the viable region of parameter space 
consistent with the LMA solar solution.  These contours are almost parallel 
to the contours of constant $\sin^2 2\theta_{13}$ also shown in 
Fig.~\ref{fig:P2-E1-albrightfig1}. This
implies a remarkable correlation between the predicted values of
$\Delta m^2_{21}$ and $\sin^2 2\theta_{13}$.  If the LMA solution is indeed 
the correct solution to explain the solar neutrino deficit observations, 
KamLAND \cite{KamLAND} is expected to provide measurements
of $\Delta m^2_{21}$ and $\sin^2 2\theta_{12}$. Hence the GUT model we are
considering will be able to give a precise prediction for $\sin^2
2\theta_{13}$ once $\Delta m^2_{21}$ and $\sin^2 2\theta_{12}$ are known.

A more detailed study of the correlations predicted by the model for points
within the allowed LMA region has been carried out in collaboration with
S. Geer \cite{albgeer}.  An additional source of CP violation in the leptonic
sector has also been studied there, whereby the two Higgs singlets breaking
lepton number and leading to $a$ and $b$ contributions are assigned a relative
phase.  Although maximal atmospheric $\nu_\mu - \nu_\tau$ mixing tends to 
favor a small leptonic $\delta_{CP}$ phase, the phase can be as large as 
$|\delta_{CP}| \aprle 50^\circ$ in the presently allowed LMA region.  
We conclude that a precise test of all the model predictions for the 
solar and atmospheric neutrino oscillations will require a Neutrino Factory.

\end{document}